\documentclass[aps,pra,amsmath,amsfonts,amssymb,twocolumn,superscriptaddress,showpacs]{revtex4}

\usepackage{amssymb}
\usepackage{graphicx,psfrag,color}
\usepackage{dcolumn}
\usepackage{bm}
\usepackage{mathbbol}

\begin{document}
\flushbottom

\title{Probabilistic quantum teleportation via thermal entanglement}

\author{Raphael Fortes}
\affiliation{Universidade Federal da Integra\c{c}\~ao Latino Americana, 85867-970,  Foz do Igua\c{c}u, Paran\'a, Brazil}
\author{Gustavo Rigolin}
\email{rigolin@ufscar.br}
\affiliation{Departamento de F\'isica, Universidade Federal de
S\~ao Carlos, 13565-905, S\~ao Carlos, S\~ao Paulo, Brazil}

\date{\today}

\begin{abstract}
We study the probabilistic (conditional) teleportation protocol 
when the entanglement needed to its implementation is given by thermal entanglement, i.e.,
when the entangled resource connecting Alice and Bob is an entangled mixed state described by
the canonical ensemble density matrix. Specifically, the entangled resource we employ here 
is given by two interacting spin-1/2 systems (two qubits) in equilibrium with a thermal reservoir at temperature $T$.
The interaction between the qubits is described by a Heisenberg-like Hamiltonian, encompassing the 
Ising, the XX, the XY, the XXX, and XXZ models, with or without external fields. For all those models we show analytically 
that the probabilistic protocol is exactly equal to the deterministic one whenever we have no external field. However,
when we turn on the field the probabilistic protocol outperforms the deterministic one in several interesting ways.
Under certain scenarios, for example, the efficiency (average fidelity) 
of the probabilistic protocol is greater than the deterministic one and
increases with increasing temperature, a
counterintuitive behavior. We also show regimes in which the probabilistic protocol operates with relatively high
success rates and, at the same time, with efficiency greater than the classical limit $2/3$, 
a threshold that cannot be surpassed by any protocol using only classical resources (no entanglement shared between 
Alice and Bob). 
The deterministic protocol's efficiency under the same
conditions is below $2/3$, highlighting that the probabilistic protocol is the only one yielding a genuine quantum teleportation.
We also show that near the quantum critical points for almost all those models the qualitative and quantitative behaviors of the efficiency change
considerably, even at finite $T$.
\end{abstract}


\maketitle

\section{Introduction} 

The quantum teleportation protocol \cite{ben93} is one of the most 
important quantum communication protocols devised so far. It was originally built \cite{ben93} to transfer 
an unknown quantum state $|\psi\rangle$, describing a qubit located in one place (Alice's), to another qubit in another place (Bob's) 
without sending the physical system originally described by $|\psi\rangle$ from Alice to Bob. 
A few years after its conception, the teleportation protocol was extended to continuous variable systems \cite{vai94,bra98} and
also the first experimental realizations were presented \cite{bow97,bos98,fur98}.  
The key resource needed to accomplish such a task without corrupting the teleported state 
is a maximally entangled pure state that Alice and Bob must share. This maximally entangled 
pure state is the ideal entangled resource through which the teleportation takes place.

Generating and preserving a maximally entangled pure state is not easy. 
Unavoidable losses, noise, and decoherence rapidly reduce its purity and entanglement. 
A workaround to bypass those problems using only local operations and classical communication 
is entanglement distillation \cite{ben96}, where several
copies of non-maximally entangled mixed states are converted into one maximally entangled pure state.
A different approach is based on the modification of the standard teleportation protocols
\cite{ben93,vai94,bra98}, adapting them to operate directly with non-maximally entangled states  
\cite{guo00,agr02,rig06,bow01,alb02,lee02,tak12,hor00,ban02,yeo08,ban12,kno14,for15,lui15}.

The modified teleportation protocols can be divided into two main groups. 
The first one contains the deterministic protocols \cite{bow01,alb02,lee02,tak12,hor00,ban02,yeo08,ban12,kno14,for15,lui15,par16},
in which there is no postselection of the measurement results obtained by Alice during the execution of the protocol. 
In other words, at the end of each run of the protocol, no matter what measurement outcome Alice obtains, 
Bob considers his qubit as a valid output of the teleportation protocol. The word ``deterministic'' means that the probability of
``success'' is one for those protocols, even if Bob's qubit at the end of the protocol is not exactly equal to the input state. 
The second group contains the probabilistic protocols, in which Alice and Bob postselect certain measurement results of Alice.  
In this scenario, Alice's measurement outcomes leading to low fidelity teleported states are discarded and thus the protocol is 
dubbed probabilistic since the chances of Alice getting the measurement results leading to high fidelity teleported states
are less than one \cite{guo00,agr02,rig06,for16}.

Most of the works dealing with probabilistic teleportation protocols employ non-maximally pure entangled states 
as the entangled resource connecting Alice and Bob \cite{guo00,agr02,rig06}. Only recently was presented a comprehensive investigation of the 
probabilistic protocol with mixed entangled states \cite{for16}. In Ref. \cite{for16} each qubit 
of a maximally entangled pure state (Bell state) was independently subjected to all possible combinations of 
the four standard types of noise one faces in the implementation of quantum communication tasks,
namely, the bit flip, the phase flip or phase dumping, the depolarizing, and the amplitude damping noise channels. 
The efficiency to teleport a qubit of each one of the 16 mixed states obtained after the action of those kinds of noise was studied. 
It was also assumed that Alice's qubit might also be acted by each one of those four types of noise, giving a total of 64 case studies. 

\begin{figure}[!ht]
\includegraphics[width=8.5cm]{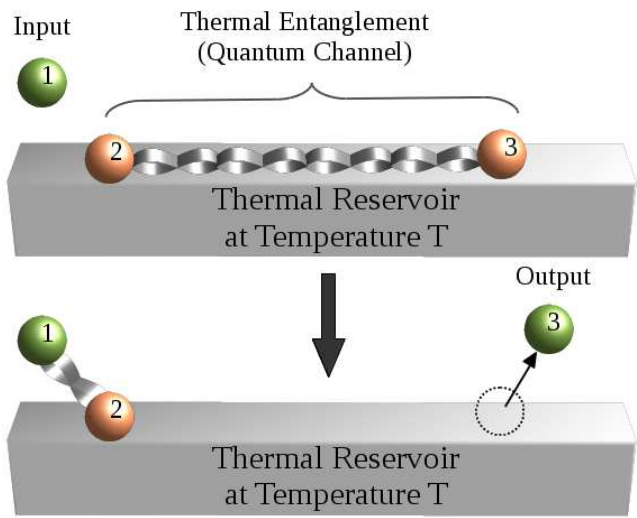}
\caption{\label{fig1}(color online) The teleportation protocol can basically be divided into four steps.
Upper panel: The first step is related to the preparation of the entangled resource (qubits 2 and 3) and the input 
(qubit 1). Here the entangled resource is described by two interacting qubits 
in thermal equilibrium with a thermal reservoir at temperature $T$.
Lower panel: The second step consists in Alice implementing a joint measurement (Bell measurement) in the input and her share
of the entangled resource (qubits 1 and 2), which become entangled. 
Step three is the broadcasting to Bob, via a classical communication channel,
of Alice's measurement result. In the fourth and last step, Bob implements a unitary operation on the output state 
(qubit 3) depending on the result of Alice's measurement. 
Note that the present analysis is particularly relevant and meaningful whenever the overall time
needed to implement all steps of the teleportation protocol is lower than the time needed 
by the whole system to be brought back to thermal equilibrium. In other words, the rate at which we
implement all steps of the teleportation protocol must be higher than the system's 
thermal relaxation rate. In the opposite scenario, however, Bob's output qubit would 
return to a thermal equilibrium state before we could access and further manipulate its content 
or even before we could finish the teleportation protocol. In this case the present analysis does 
not apply.
}
\end{figure}

In this manuscript our goal is to study a different yet important noise scenario. We now consider that
the two qubits shared between Alice and Bob can interact and that they are in thermal equilibrium with a thermal reservoir 
at temperature $T$ (see Fig. \ref{fig1}). This scenario naturally appears in a possible implementation of a quantum computer 
based on solid state devices, where quantum information needs to be transferred (teleported) from one location to  
another inside a quantum chip and $T$ is the temperature under which the quantum computer operates.

The quantum state of a two-qubit system in equilibrium with a thermal reservoir is described by the
canonical ensemble density matrix and whenever entanglement is present between the two qubits it is usually called 
\textit{thermal entanglement} \cite{Ved01,Kam02,Rig03,Ami07,reviews,Wer10,werPRL,werPRA,werIJMPB}. In this manuscript
we model the interaction between the qubits of the entangled resource via the Heisenberg Hamiltonian, either without or with an external magnetic
field. The external magnetic field gives an important extra control parameter that can be tuned to maximize the teleportation efficiency.
For several combinations of the coupling constants and external field in the Heisenberg Hamiltonian, we obtain counter-intuitive situations where
an increase of the temperature leads to a better teleportation. Also, we show that 
there are cases where the probabilistic protocol beats the deterministic one in a very
important way, already seen in the noise models of Ref. \cite{for16}: 
we prove that for some set of coupling constants in the Heisenberg model, the probabilistic protocol 
is the only one leading to a genuine quantum teleportation. This is true because 
the deterministic protocol under the same conditions cannot overcome the efficiency (average fidelity) of an all-classical protocol, where
no entanglement is used to teleport the qubit. The probabilistic protocol, on the other hand, has an efficiency that cannot
be achieved by the all-classical protocol. We also investigate how the 
efficiencies of the probabilistic and deterministic protocols are affected in the vicinity of the quantum
critical points for the models we study here. We noted non-trivial qualitative and quantitative changes in the
behavior of the efficiencies near the critical points, even at finite $T$.

\section{The mathematical tools}
\label{secI}

Since the entangled resource in the present manuscript is not a pure state, we have to recast
the original teleportation protocol using the language of density matrices. This was 
done for the deterministic protocol in Ref. \cite{for15} and for the probabilistic protocol in Ref. \cite{for16}.
In Secs. \ref{densitymatrix} and \ref{fidelity} we review the main ideas and results of those references that are 
needed here. We follow closely the notation and style of Refs. \cite{for15,for16}. In Sec. \ref{heisenberg}
we present the Heisenberg model, preparing the ground for Sec. \ref{results}, where we show the main results of this manuscript.

\subsection{Recasting the teleportation protocol in the density matrix formalism}
\label{densitymatrix}

The input qubit that is teleported from Alice to Bob is assumed a pure state and is given by  
$|\psi\rangle_{in}= a|0\rangle + b|1\rangle$, with $|a|^2+|b|^2=1$. Its density matrix is  
\begin{equation}
\rho_{in} = |\psi\rangle_{in}\;_{in}\langle\psi | =
\left(
\begin{array}{cc}
|a|^2 & ab^* \\
a^*b & |b|^2
\end{array}
\right),
\label{step0}
\end{equation}
where $*$ is complex conjugation and the subscript $in$ means ``input''.  
The entangled state shared between Alice and Bob (the quantum communication channel) is described by the canonical ensemble density matrix, 
\begin{equation}
\rho_{ch} = \frac{e^{-\frac{H}{kT}}}{Z}=\frac{e^{-\beta H}}{Z},
\label{canonical}
\end{equation}
where $Z=\mbox{Tr}(e^{-H/kT})$ is the partition function, ``Tr'' is the trace operation, $k$ is the Boltzmann constant,
and $ch$ means the quantum communication ``channel''. The Hamiltonian $H$ is given by the Heisenberg model as explained in Sec. \ref{heisenberg}.
Note that the expression ``communication channel'' refers to any physical apparatus, device or system whereby Alice and Bob 
may send either classical or quantum information. In the former case we call it a classical communication channel and
in the latter a quantum communication channel. Throughout this text the words entangled resource
and quantum communication channel are synonyms.

At this stage, the total state describing all qubits is 
\begin{equation}
\rho = \rho_{in} \otimes \rho_{ch}.
\label{step1}
\end{equation}

The teleportation protocol proceeds as follows:
\begin{enumerate}
\item[(i)] Alice implements a Bell state measurement onto qubits 1 and 2. 
\item[(ii)] Alice broadcasts her measurement result to Bob using a classical communication channel.
\item[(iii)] Bob uses the information received in step (ii) to choose the right unitary operation 
to be applied on his state (qubit 3).
\end{enumerate}

If Alice and Bob shared a maximally entangled pure state (Bell state), at the end of step (iii) Bob's qubit
would be exactly described by $\rho_{in}$. In any realistic scenario this is not the case and we invariably have a mixed state 
describing the quantum communication channel, leading to a non-perfect teleportation.

The projectors describing Alice's measurement on the input qubit and her qubit of the entangled resource are
\begin{equation}
P_j^\varphi=|B_j^\varphi\rangle\langle B_j^\varphi|, \hspace{.5cm} j=1,2,3,4,
\label{Bj}
\end{equation}
with
\begin{eqnarray}
|B_1^\varphi\rangle&=&\cos\varphi|00\rangle+\sin\varphi|11\rangle, \label{B1} \\
|B_2^\varphi\rangle&=&\sin\varphi|00\rangle-\cos\varphi|11\rangle, \\
|B_3^\varphi\rangle&=&\cos\varphi|01\rangle+\sin\varphi|10\rangle, \label{B3} \\
|B_4^\varphi\rangle&=&\sin\varphi|01\rangle-\cos\varphi|10\rangle. \label{B4}
\end{eqnarray}
In the standard protocol $\varphi=\pi/4$ and $|B_j\rangle$, $j=1,2,3,4$, are respectively 
the Bell states $|\Phi^+\rangle, |\Phi^-\rangle, |\Psi^+\rangle$, and $|\Psi^-\rangle$  \cite{ben93}. Here $\varphi$
is a free parameter chosen by Alice to maximize the efficiency of the probabilistic 
teleportation. 

Alice's probability to measure a given generalized Bell state is
\begin{equation}
Q_j(|\psi\rangle_{in}) = \mbox{Tr}[{P_j^\varphi \rho}]
\label{prob}
\end{equation}
and at the end of step (iii) Bob's state is 
\begin{equation}
\rho_{_{B_j}}=   \frac{U_j\mbox{Tr}_{12}[P_j^\varphi \rho P_j^\varphi]U_j^\dagger}{Q_j(|\psi\rangle_{in})}.
\label{twelve}
\end{equation}
Here 
$\mbox{Tr}_{12}$ is the partial trace on the first two qubits (Alice's qubits). 
%
%
We make it explicit the dependence of $Q_j$ on the input state
$|\psi\rangle_{in}$ since for mixed state entangled resources, or non-maximally entangled ones,
the probability depends on the input state \cite{guo00,agr02,rig06,for15,for16}. 

In the standard teleportation protocol the unitary transformation that Bob implements on his qubit 
depends not only on Alice's measurement outcome but also on the entangled resource \cite{ben93}. For example, if $\rho_{ch}$ is the Bell state
$|\Phi^+\rangle=(|00\rangle+|11\rangle)/\sqrt{2}$ we have $U_1=\mathbb{1}$, $U_2=\sigma_z, U_3=\sigma_x$, and $U_4=\sigma_z\sigma_x$, 
with $\mathbb{1}$ being the identity matrix and $\sigma_z$ and $\sigma_x$ the standard Pauli matrices. 
For the other three Bell states,  $|\Phi^-\rangle, |\Psi^+\rangle$, and $|\Psi^-\rangle$, we have respectively
$\{U_1,U_2,U_3,U_4\}=\{\sigma_z,\mathbb{1},\sigma_z\sigma_x,\sigma_x\},\{\sigma_x,\sigma_z\sigma_x,\mathbb{1},\sigma_z\}$, and 
$\{\sigma_z\sigma_x,\sigma_x,\sigma_z,\mathbb{1}\}$. With that in mind, when we search for the optimal
settings leading to the greatest efficiency for the teleportation protocol, we will also 
let $U_j$ run over its possible four values: $\mathbb{1},\sigma_z,\sigma_x,\sigma_z\sigma_x$.

Before we proceed it is worth better explaining what we mean by optimal settings or 
an optimal protocol. In the following we will be looking for the optimal protocols for
several entangled resources shared between Alice and Bob. Our search for the optimal 
protocols, the ones leading to the greatest efficiencies (average fidelities), will be restricted
to projective measurements that Alice might implement on her qubits and Bob 
will be restricted to act on his qubit using only Pauli matrices, as explained in the
previous paragraph. We decided to work only with projective measurements and with
unitary operations given by Pauli matrices because those are the resources employed in the original
protocol and readily implementable with current technology. It lies beyond the scope and aim of this
work to deal with more general types of measurements and more general unitary operations.

\subsection{Success rate and efficiency of the probabilistic teleportation}
\label{fidelity}

Since the chance $Q_j$ of Alice measuring the generalized Bell state $|B_j^\varphi\rangle$ 
when she shares with Bob a non-maximally entangled resource 
depends on the input state $|\psi\rangle_{in}$ \cite{guo00,agr02,rig06,for15,for16}, we assume $|\psi\rangle_{in}$ is given by a uniform probability 
distribution,
\begin{equation}
P_X(x)=\mathcal{P}(|\psi\rangle_{in}).
\end{equation}
Here $X$ denotes a continuous random variable whose values $x$ are all possible pure qubits that
together define the sample space $\Omega$. Averaging over this distribution we obtain input-state-independent results
for the relevant quantities needed to study the efficiency of the teleportation protocol. 
The probability distribution $P_X(x)$ is normalized as follows,  
\begin{equation}
\int_{\Omega} P_X(x) dx=\int_{\Omega}\mathcal{P}(|\psi\rangle_{in})d|\psi\rangle_{in} = 1,
\label{normalization1}
\end{equation}
where $P_X(x)$ is constant for all $x$. 

If we write an arbitrary qubit as
\begin{equation}
|\psi\rangle = \alpha|0\rangle + \delta e^{i\gamma}|1\rangle,
\label{relative}
\end{equation}
where $\alpha\geq 0$, $\delta\geq 0$, $\alpha^2+\delta^2=1$, and $0\leq\gamma\leq 2\pi$ are real numbers,
it is not difficult to see that we can select $\alpha^2$ and $\gamma$ as independent variables. 
Thus $\mathcal{P}(|\psi\rangle_{in})=\mathcal{P}(\alpha^2,\gamma)$ and 
 Eq.~(\ref{normalization1}) reads
\begin{equation}
\int_{0}^{2\pi}\int_{0}^{1}\mathcal{P}(\alpha^2,\gamma)\mathrm{d}\alpha^2\mathrm{d}\gamma =1,
\label{normP}
\end{equation}
where 
\begin{equation}
\mathcal{P}(\alpha^2,\gamma)=\frac{1}{2\pi}
\label{uniform}
\end{equation}
for a uniform distribution.

It is worth mentioning that the described averaging over pure states does not correspond to a
uniform distribution on the Bloch sphere. The distribution as given here states that  the relative phase between the states 
$|0\rangle$ and $|1\rangle$ is completely random as well as the probability weight of the state $|0\rangle$  (or $|1\rangle$) 
in the superposition of $|0\rangle$ and $|1\rangle$. Nevertheless, this distribution is easy to implement in 
the laboratory and from a mathematical and operational point of view, we have observed that it 
simplifies the calculations of the average fidelities, in particular for the probabilistic protocols. 

There is also a discrete variable $J$ with values $j=1,2,3,4$ (or $j=\Phi^+,\Phi^-,\Psi^+,\Psi^-$) representing 
the generalized Bell states $|B_j^\varphi\rangle$. Thus, the probability to measure $|B_j^\varphi\rangle$
is denoted by $P_J(j)$. The conditional probability $P_{J|X}(j|x)$ gives Alice's chance of measuring the Bell state $j$ if
the input state to be teleported is $x$ and is given by Eq. (\ref{prob}),
\begin{equation}
P_{J|X}(j|x) = Q_j(|\psi\rangle_{in}).
\end{equation}

The joint probability distribution $P_{XJ}(x,j)=P_{JX}(j,x)$ can be obtained if we 
use the definition of the conditional probability, 
\begin{equation}
P_{XJ}(x,j)=P_{X}(x)P_{J|X}(j|x)=\mathcal{P}(|\psi\rangle_{in})Q_j(|\psi\rangle_{in}),
\label{relationP}
\end{equation}
which subsequently allows us to compute the marginal distribution $P_J(j)=\int_\Omega P_{XJ}(x,j) dx$,
\begin{equation}
P_{J}(j) = \int_\Omega \mathcal{P}(|\psi\rangle_{in})Q_j(|\psi\rangle_{in}) d|\psi\rangle_{in}.
\label{marginalJ}
\end{equation}
And if we use Eq.~(\ref{relationP}) exchanging the roles of $X$
with $J$ and Eq.~(\ref{marginalJ}) we arrive at
\begin{eqnarray}
P_{X|J}(x|j) &=& 
\frac{P_{XJ}(x,j)}{P_{J}(j)} 
\nonumber \\
&=&  
\frac{\mathcal{P}(|\psi\rangle_{in})Q_j(|\psi\rangle_{in})}
{\int_\Omega \mathcal{P}(|\psi\rangle_{in})Q_j(|\psi\rangle_{in}) d|\psi\rangle_{in}}.
\label{finalP}
\end{eqnarray}
These last two expressions, Eqs. (\ref{marginalJ}) and (\ref{finalP}), are the probability distributions needed to 
quantitatively study the probabilistic teleportation protocol. 

We can better appreciate the last statement remembering the meaning of $P_{J}(j)$ and $P_{X|J}(x|j)$. 
Noting that $P_{J}(j)$ gives the chance of Alice measuring the generalized Bell state $|B_j^\varphi\rangle$  
when the distribution for the input states is $\mathcal{P}(|\psi\rangle_{in})$, it is straightforward to see that $P_{J}(j)$ 
is the average probability of measuring $|B_j^\varphi\rangle$,
\begin{equation}
\overline{Q}_j= P_{J}(j) = \int_\Omega \mathcal{P}(|\psi\rangle_{in})Q_j(|\psi\rangle_{in}) d|\psi\rangle_{in}.
\label{averageQ}
\end{equation}
$\overline{Q}_j$ does not dependent on $|\psi\rangle_{in}$ and is called the probability
of success or the success rate of the probabilistic teleportation protocol if we postselect the measurement result $j$ \cite{for16}.


To quantify how similar to the input state is the output after one run of the protocol we employ the fidelity \cite{uhl76},
which for a pure input state is
\begin{equation}
F_j(|\psi \rangle_{in}) = \mbox{Tr}[\rho_{in}\rho_{_{B_j}}]=\,_{in}\langle \psi | \rho_{_{B_j}} | \psi \rangle_{in},
\label{Fj}
\end{equation}
with $\rho_{_{B_j}}$, Eq.~(\ref{twelve}), being the output state with Bob after the teleportation protocol ends. 
For a perfect teleportation $F_j=1$ (its maximal value) and $F_j=0$ (its minimal value) 
when the output is orthogonal to the input state.

Looking at Eq.~(\ref{Fj}) we see that in general $F_j$ depends on $|\psi\rangle_{in}$ and by averaging over all possible input states 
we get an input-state-independent quantification for the efficiency of the protocol \cite{for16}. Since we are interested in a postselected
measurement result $j$, the distribution of input states $|\psi\rangle_{in}$ in this situation is $P_{X|J}(x|j)$, 
Eq.~(\ref{finalP}), which leads to the following average fidelity,
\begin{eqnarray}
\overline{F}_j &=& \int_{\Omega}F_j(x)P_{X|J}(x|j)dx \nonumber \\
&=&
\frac{\int_\Omega F_j(| \psi \rangle_{in})\mathcal{P}(|\psi\rangle_{in})Q_j(|\psi\rangle_{in})d|\psi\rangle_{in}}
{\int_\Omega \mathcal{P}(|\psi\rangle_{in})Q_j(|\psi\rangle_{in}) d|\psi\rangle_{in}}.
\label{FinalFj} 
\end{eqnarray}
This is what we call the efficiency of the probabilistic teleportation protocol if we postselect the 
measurement result $j$ \cite{for16}. 
If all measurement results are accepted, i.e., no postselection is made, 
we get back the efficiency of the deterministic protocol \cite{for15,for16}, 
\begin{equation}
\langle \overline{F} \rangle = \sum_{j=1}^{4}P_J(j)\overline{F}_j = 
\int_\Omega \overline{F}(|\psi\rangle_{in}) \mathcal{P}(|\psi\rangle_{in})d|\psi\rangle_{in},
\label{deterministicF}
\end{equation}
where $\overline{F}(|\psi\rangle_{in})= \sum_j^4Q_j(|\psi\rangle_{in})F_j(| \psi \rangle_{in})$. 

Following the strategy of Ref. \cite{for16}, we want to maximize Eq.~(\ref{FinalFj}) over the set of free parameters
present in the probabilistic protocol. In particular, we want to get scenarios in which $\overline{F}_j > \langle\overline{F}\rangle$, 
where $\langle\overline{F}\rangle$ is the optimal efficiency of the deterministic teleportation protocol.

\subsection{The Heisenberg model}
\label{heisenberg}

The Hamiltonian describing the Heisenberg model for a spin-1/2 chain of two qubits is
\begin{equation}
H = j_x \sigma_x^{(2)} \sigma_x^{(3)}+j_y \sigma_y^{(2)} \sigma_y^{(3)}+j_z \sigma_z^{(2)} \sigma_z^{(3)}
+h_a \sigma_z^{(2)}+h_b \sigma_z^{(3)},
\label{hamiltonian}
\end{equation}
where $\sigma_j^{(2)} \sigma_j^{(3)}=\sigma_j^{(2)} \otimes \sigma_j^{(3)}$, with  
the superscripts $(2)$ and $(3)$ representing qubits 2 (with Alice) and 3 (with Bob) of the quantum communication
channel (see Fig. \ref{fig1}). In Eq.~(\ref{hamiltonian}), $\sigma_j$, $j=x,y,z$, are the standard Pauli matrices
such that $\sigma_z|0\rangle=|0\rangle$ and $\sigma_z|1\rangle=-|1\rangle$, 
$\sigma_x|0\rangle=|1\rangle$ and $\sigma_x|1\rangle=|0\rangle$, and 
$\sigma_y|0\rangle=i|1\rangle$ and $\sigma_y|1\rangle=-i|0\rangle$, with $i$ being the imaginary unity.
Furthermore, $j_x,j_y,j_z,h_a,h_b$ are real numbers with the former three representing the coupling
constants between the qubits and the latter two denoting external magnetic fields applied respectively
on qubits 2 and 3 along the $z$ direction.

Inserting Eq.~(\ref{hamiltonian}) into Eq.~(\ref{canonical}) we get the canonical ensemble
density matrix describing the quantum communication channel $\rho_{ch}$, which together with Eq.~(\ref{step0}) 
allows us to compute the total state $\rho$ initially describing all three qubits employed in the
teleportation protocol (see Eq.~(\ref{step1})). Using $\rho$ we can evaluate Eq.~(\ref{prob}) 
and insert it along with Eq.~(\ref{uniform}) into
Eq.~(\ref{averageQ}) to obtain the four success rates $\overline{Q}_j$, 
each of which is associated with the  average probability of measuring 
the generalized Bell state $|B_j^\varphi\rangle$. Those success rates
can be written as follows,
\begin{eqnarray}
\overline{Q}_1=\overline{Q}_4 &=& q(\varphi), \\
\overline{Q}_2= \overline{Q}_3 &=& q(\pi/2\pm\varphi), 
\end{eqnarray}
where
\begin{equation}
q(\varphi) = \frac{1}{4}-\frac{\cos (2 \varphi) \left[\eta \Delta_h   \sinh (\beta  \chi )+\chi  \Sigma_h 
e^{2 \beta  j_z} \sinh (\beta  \eta )\right]}{4 \chi  \eta  \left[\cosh (\beta  \chi )+e^{2 \beta j_z} \cosh (\beta  \eta )\right]}.
\label{q}
\end{equation}
In Eq.~(\ref{q}), $\beta=1/kT$ and $\varphi$ were already defined in Eqs.~(\ref{canonical}) and (\ref{Bj}), respectively, 
while the other quantities are given as follows,
\begin{eqnarray}
\eta = \sqrt{\Delta_j^2+\Sigma_h^2}, & \Delta_j = j_x - j_y, & \Sigma_h = h_a + h_b, \label{eta}\\
\chi = \sqrt{\Delta_h^2+\Sigma_j^2}, & \Delta_h = h_a - h_b, & \Sigma_j = j_x + j_y. \label{chi}
\end{eqnarray}

We now turn our attention to the efficiency of the teleportation protocol (average fidelities). 
Before we proceed it is important to recall that the unitary operation $U_j$ that Bob must implement on
his qubit at the end of the protocol depends, in addition to Alice's measurement result, 
on which quantum communication channel (entangled state) she shares with Bob. In the original protocol \cite{ben93}, 
for each one of the four possible Bell states (maximally entangled pure states) that Alice and Bob might share, 
we can associate a set $S$ containing four $U_j$. Each member of $S$ 
corresponds to the unitary operation that Bob needs to implement on his qubit 
according to Alice's measurement result (see Sec.~\ref{densitymatrix}).

Here we deal with a mixed state entangled resource which, similarly to any two-qubit state, can be written as
$\rho_{ch}=p_{\!_{\Phi^+}}|\Phi^+\rangle\langle\Phi^+|+p_{\!_{\Phi^-}}|\Phi^-\rangle\langle\Phi^-|
+p_{\!_{\Psi^+}}|\Psi^+\rangle\langle\Psi^+|+p_{\!_{\Psi^-}}|\Psi^-\rangle\langle\Psi^-|$ $+$ non-diagonal terms.
We are employing the Bell states as a basis to expand $\rho_{ch}$ and thus $p_j$, $j=\Phi^+, \Phi^-,\Psi^+,\Psi^-$, 
are the probabilities of projecting $\rho_{ch}$ onto the respective Bell states.
Depending on the parameters of Eq.~(\ref{hamiltonian}), one (or more) $p_j$ dominates and it is expected that 
the set $S$ associated with the corresponding Bell state will yield the best efficiency for the 
teleportation protocol. Therefore, in our search for the optimal protocol, 
we compute the efficiencies of the probabilistic and deterministic protocols, 
Eqs.~(\ref{FinalFj}) and (\ref{deterministicF}), using the four possible sets $S$. In the end,
i.e., after we optimize all expressions with respect to the free parameters of the protocol, 
we pick out of all possibilities the one giving the greatest efficiency.

\subsubsection{The deterministic protocol}

Let us begin analyzing the efficiency for the deterministic
protocol, Eq.~(\ref{deterministicF}), where we append a superscript to $\langle \overline{F} \rangle$
to remind us of which set $S=\{U_1,U_2,U_3,U_4\}$ of unitary operations we employ in the
calculation of $\langle \overline{F} \rangle$. For example,  
$\langle \overline{F} \rangle^{\!^{\Phi^+}}$ means that we use the set $S$ associated to 
the case where the entangled resource is the Bell state $|\Phi^+\rangle$ (see Sec.~\ref{densitymatrix}).
Using Eqs.~(\ref{prob}), (\ref{twelve}), (\ref{uniform}), and (\ref{Fj}) in Eq.~(\ref{deterministicF}) we
get
\begin{eqnarray}
\langle \overline{F} \rangle^{\!^{\Phi^+}} &=& f^{\!_{\Phi}}(\varphi), \\
\langle \overline{F} \rangle^{\!^{\Phi^-}} &=& f^{\!_{\Phi}}(-\varphi),\\
\langle \overline{F} \rangle^{\!^{\Psi^+}} &=& f^{\!_{\Psi}}(\varphi), \\
\langle \overline{F} \rangle^{\!^{\Psi^-}} &=& f^{\!_{\Psi}}(-\varphi), 
\end{eqnarray}
where
\begin{eqnarray}
f^{\!_{\Phi}}(\varphi) &=& \frac{1}{3} + \frac{\chi  \cosh (\beta  \chi )-\Sigma_j \sin (2 \varphi) \sinh (\beta  \chi )}
{3 \chi  \left[\cosh (\beta  \chi )+e^{2 \beta  j_z} \cosh (\beta  \eta )\right]}, \label{fphi}\\
f^{\!_{\Psi}}(\varphi) &=& \frac{1}{3} + \frac{\eta  \cosh (\beta  \eta )-\Delta_j \sin (2 \varphi) \sinh (\beta  \eta )}
{3 \eta  \left[e^{-2 \beta  j_z} \cosh (\beta  \chi )+\cosh (\beta  \eta )\right]} \label{fpsi}.
\end{eqnarray}

Looking at Eqs.~(\ref{fphi}) and (\ref{fpsi}), and noting that $\beta$, $\chi$, and $\eta$ are positive 
quantities, we easily see that the optimal expressions are obtained by setting $\varphi=\pm \pi/4$. In other words, the measurement
basis Alice must employ is the standard Bell basis. More specifically, we must choose $\varphi$ such that 
$-\Sigma_j \sin (2 \varphi)=|\Sigma_j|$ and $-\Delta_j \sin (2 \varphi) = |\Delta_j|$.
If $\Sigma_j < 0$ we choose $\varphi=\pi/4$ and when $\Sigma_j> 0$ we set $\varphi=-\pi/4$ (or $5\pi/4$). 
A similar analysis applies to $\Delta_j$. 
Therefore, the optimal average fidelities for each set $S$ 
are
\begin{eqnarray}
\langle \overline{F} \rangle^{\!^{\Phi^+}}_{\!_{opt}} = &  \langle \overline{F} \rangle^{\!^{\Phi^-}}_{\!_{opt}}  & = f^{\!_{\Phi}}_{\!_{opt}},
\label{fdetoptphi}\\
\langle \overline{F} \rangle^{\!^{\Psi^+}}_{\!_{opt}} = & \langle \overline{F} \rangle^{\!^{\Psi^-}}_{\!_{opt}} & = f^{\!_{\Psi}}_{\!_{opt}}, 
\label{fdetoptpsi}
\end{eqnarray}
where
\begin{eqnarray}
f^{\!_{\Phi}}_{\!_{opt}}  &=& \frac{1}{3} + \frac{\chi  \cosh (\beta  \chi )+|\Sigma_j|  \sinh (\beta  \chi )}
{3 \chi  \left[\cosh (\beta  \chi )+e^{2 \beta  j_z} \cosh (\beta  \eta )\right]}, \label{fphiopt}\\
f^{\!_{\Psi}}_{\!_{opt}} &=& \frac{1}{3} + \frac{\eta  \cosh (\beta  \eta )+|\Delta_j|  \sinh (\beta  \eta )}
{3 \eta  \left[e^{-2 \beta  j_z} \cosh (\beta  \chi )+\cosh (\beta  \eta )\right]} \label{fpsiopt}.
\end{eqnarray}

Finally, the optimal efficiency for the deterministic teleportation protocol is given by
\begin{equation}
\langle \overline{F} \rangle_{\!_{opt}} =\max\{f^{\!_{\Phi}}_{\!_{opt}} ,f^{\!_{\Psi}}_{\!_{opt}} \}.
\label{optimalFdeterministic}
\end{equation}
Equation (\ref{optimalFdeterministic}) is the benchmark we want to surpass using the probabilistic protocol.

\subsubsection{The probabilistic protocol}

Following the superscript notation just introduced in the preceding analysis, 
we now need to evaluate $\overline{F}_j^{\;\epsilon}$, Eq.~(\ref{FinalFj}), for $j=1,2,3,4$ and $\epsilon=\Phi^+,\Phi^-,\Psi^+,\Psi^-$. 
Here each $j$ represents one of the four possible measurement outcomes of Alice, i.e., it denotes which generalized Bell state $|B^\varphi_j\rangle$
she measured, and $\epsilon$ represents which set of unitary operations $S$ Bob uses to properly correct his
qubit, where each element of the set corresponds to a given measurement result of Alice.  
For instance, $\overline{F}_1^{\,_{\Phi^+}}$ means that Alice and Bob are working with the postselected
measurement outcome $|B^\varphi_1\rangle$, discarding the other three possible measurement results,
and Bob's unitary operation for all valid runs of the protocol is always $\mathbb{1}$ (the respective $U_1$ associated with $\epsilon=\Phi^+$). 
In Table \ref{table} we list all $16$ possibilities.
\begin{table}[!ht]
\caption{\label{table} In the table below we list to each $\overline{F}_j^{\,_{\epsilon}}$ the corresponding
Alice's measurement outcome $|B_j^\varphi\rangle$ and the respective unitary operation Bob implements on his qubit.}
\begin{ruledtabular}
\begin{tabular}{ll}
\begin{tabular}{lll}
$\overline{F}_1^{\,_{\Phi^+}} \longrightarrow$ & $|B^\varphi_1\rangle\longrightarrow$ & $\mathbb{1}$ \\
$\overline{F}_2^{\,_{\Phi^+}}\longrightarrow$ & $|B^\varphi_2\rangle\longrightarrow$ & $\sigma_z$ \\
$\overline{F}_3^{\,_{\Phi^+}}\longrightarrow$ & $|B^\varphi_3\rangle\longrightarrow$ & $\sigma_x$ \\
$\overline{F}_4^{\,_{\Phi^+}}\longrightarrow$ & $|B^\varphi_4\rangle\longrightarrow$ & $\sigma_z\sigma_x$
\end{tabular}
&
\begin{tabular}{lll}
$\overline{F}_1^{\,_{\Phi^-}}\longrightarrow$ & $|B^\varphi_1\rangle\longrightarrow$ & $\sigma_z$ \\
$\overline{F}_2^{\,_{\Phi^-}}\longrightarrow$ & $|B^\varphi_2\rangle\longrightarrow$ & $\mathbb{1}$ \\
$\overline{F}_3^{\,_{\Phi^-}}\longrightarrow$ & $|B^\varphi_3\rangle\longrightarrow$ & $\sigma_z\sigma_x$ \\
$\overline{F}_4^{\,_{\Phi^-}}\longrightarrow$ & $|B^\varphi_4\rangle\longrightarrow$ & $\sigma_x$ 
\end{tabular}
\\ \hline
\begin{tabular}{lll}
$\overline{F}_1^{\,_{\Psi^+}}\longrightarrow$ & $|B^\varphi_1\rangle\longrightarrow$ & $\sigma_x$ \\
$\overline{F}_2^{\,_{\Psi^+}}\longrightarrow$ & $|B^\varphi_2\rangle\longrightarrow$ & $\sigma_z\sigma_x$ \\
$\overline{F}_3^{\,_{\Psi^+}}\longrightarrow$ & $|B^\varphi_3\rangle\longrightarrow$ & $\mathbb{1}$\\
$\overline{F}_4^{\,_{\Psi^+}}\longrightarrow$ & $|B^\varphi_4\rangle\longrightarrow$ & $\sigma_z$ \\
\end{tabular}
&
\begin{tabular}{lll}
$\overline{F}_1^{\,_{\Psi^-}}\longrightarrow$ & $|B^\varphi_1\rangle\longrightarrow$ & $\sigma_z\sigma_x$ \\
$\overline{F}_2^{\,_{\Psi^-}}\longrightarrow$ & $|B^\varphi_2\rangle\longrightarrow$ & $\sigma_x$ \\
$\overline{F}_3^{\,_{\Psi^-}}\longrightarrow$ & $|B^\varphi_3\rangle\longrightarrow$ & $\sigma_z$\\
$\overline{F}_4^{\,_{\Psi^-}}\longrightarrow$ & $|B^\varphi_4\rangle\longrightarrow$ & $\mathbb{1}$ 
\end{tabular}
\end{tabular}
\end{ruledtabular}
\end{table}
 
Inserting Eqs.~(\ref{prob}), (\ref{uniform}), and (\ref{Fj}) into (\ref{FinalFj}), and using the proper
unitary operation $U_j$ (see Table \ref{table}) to compute $\rho_{\!_{B_j}}$, Eq.~(\ref{twelve}), we get
\begin{eqnarray}
\overline{F}_1^{\,_{\Phi^+}}=\overline{F}_4^{\,_{\Phi^+}}&=&g^{\,_{\Phi}}(\varphi), \label{1of8}\\
\overline{F}_2^{\,_{\Phi^+}}=\overline{F}_3^{\,_{\Phi^+}}&=&g^{\,_{\Phi}}(\pi/2-\varphi), \\
\overline{F}_1^{\,_{\Phi^-}}=\overline{F}_4^{\,_{\Phi^-}}&=&g^{\,_{\Phi}}(-\varphi),\\
\overline{F}_2^{\,_{\Phi^-}}=\overline{F}_3^{\,_{\Phi^-}}&=&g^{\,_{\Phi}}(\pi/2+\varphi), \\
\overline{F}_1^{\,_{\Psi^+}}=\overline{F}_4^{\,_{\Psi^+}}&=&g^{\,_{\Psi}}(\varphi), \\
\overline{F}_2^{\,_{\Psi^+}}=\overline{F}_3^{\,_{\Psi^+}}&=&g^{\,_{\Psi}}(\pi/2-\varphi), \\
\overline{F}_1^{\,_{\Psi^-}}=\overline{F}_4^{\,_{\Psi^-}}&=&g^{\,_{\Psi}}(-\varphi), \\
\overline{F}_2^{\,_{\Psi^-}}=\overline{F}_3^{\,_{\Psi^-}}&=&g^{\,_{\Psi}}(\pi/2+\varphi), \label{8of8}
\end{eqnarray} 
where
\begin{widetext}
\begin{eqnarray}
g^{\,_{\Phi}}(\varphi) &=& \frac{1}{3} + \frac{\eta  \{\chi  \cosh (\beta  \chi )-\sinh (\beta  \chi ) [\Delta_h \cos (2 \varphi)+\Sigma_j \sin (2 \varphi)]\}}
{3 \left\{\eta\chi  \left[\cosh (\beta  \chi )+e^{2 \beta  j_z} \cosh (\beta  \eta )\right]-\cos (2 \varphi) \left[\eta\Delta_h  \sinh (\beta  \chi )
+\chi  \Sigma_h e^{2 \beta  j_z} \sinh (\beta  \eta )\right]\right\}}, \label{gphi}\\
g^{\,_{\Psi}}(\varphi) &=&\frac{1}{3}+\frac{\chi  \{\eta  \cosh (\beta  \eta )-\sinh (\beta  \eta ) [\Delta_j \sin (2 \varphi)+\Sigma_h \cos (2 \varphi)]\}}
{3\{ \eta  \chi  \left[e^{-2 \beta  j_z} \cosh (\beta  \chi )+ \cosh (\beta  \eta )\right]- \cos (2 \varphi) \left[ \eta \Delta_h e^{-2 \beta  j_z} \sinh (\beta  \chi )
+\chi  \Sigma_h  \sinh (\beta  \eta )\right]\}}. \label{gpsi}
\end{eqnarray}
\end{widetext}

The first important thing worth noting if we look at 
Eqs.~(\ref{gphi}) and (\ref{gpsi}) is the fact that $\varphi=\pm \pi/4$ (or $\varphi = \pm 3\pi/4$)  are
not in general the optimal settings. In other words, the optimal measurement basis are not formed by the standard 
maximally entangled Bell states. Indeed, whenever an external magnetic field is present,
either $\Delta_h$ or $\Sigma_h$ (or both) is not zero. This leads to the presence of the $\cos(2\varphi)$
terms, in addition to the $\sin(2\varphi)$ terms, in Eqs.~(\ref{gphi}) and (\ref{gpsi}). 
The optimal $\varphi$ in this case can be found by solving
the equations $dg^{\epsilon}/d\varphi=0$, $\epsilon = \Phi, \Psi$, and then selecting the 
$g^{\epsilon}$ giving the greatest efficiency. 

Second, comparing Eqs.~(\ref{gphi}) and (\ref{gpsi}) with (\ref{fphi}) and (\ref{fpsi}), it is not difficult to see that
\begin{eqnarray}
g^{\epsilon}(\varphi) = f^{\epsilon}(\varphi), & \mbox{if} & \Delta_h = \Sigma_h = 0.   
\end{eqnarray}
This means that if we have no external fields ($\Delta_h = \Sigma_h = 0$), the probabilistic teleportation 
protocol gives exactly the same efficiencies of the deterministic protocol. We thus arrive at the important
conclusion that the probabilistic protocol can only beat the deterministic one if
external magnetic fields are turned on. 

There is another interesting feature of the present probabilistic protocol. 
Looking at Eqs.~(\ref{1of8})-(\ref{8of8}) we see that we always have 
$\overline{F}_1^{{\epsilon}}=\overline{F}_4^{{\epsilon}}$ and
$\overline{F}_2^{{\epsilon}}=\overline{F}_3^{{\epsilon}}$, which implies that
$\overline{F}_1^{{\epsilon}}$ and $\overline{F}_4^{{\epsilon}}$, 
and equivalently $\overline{F}_2^{{\epsilon}}$ and $\overline{F}_3^{{\epsilon}}$,
share the same optimal $\varphi$. This property enhances the effective success rate of
the probabilistic protocol since two out of four possible measurement results of
Alice give the same optimal efficiency with the same optimal settings. 
Thus, instead of postselecting only one outcome, Alice and Bob can postselect two
measurement outcomes, increasing the success rate to twice the
value given in Eq.~(\ref{q}),
\begin{eqnarray}
\overline{Q}_{1,4} & = & 2 q(\varphi), \\
\overline{Q}_{2,3} & = & 2 q(\varphi\pm \pi/2). 
\end{eqnarray}

Finally, putting together all the pieces of information in the last paragraphs, and noting that 
in Eqs.~(\ref{gphi}) and (\ref{gpsi}) the arguments of all sines and cosines are given by $2\varphi$, we can
obtain the optimal efficiency for the probabilistic protocol by solving the following maximization problem,
\begin{equation}
 \overline{F}_{\!_{opt}} =\max_{\varphi\in [0,\pi]}\{g^{\!_{\Phi}}(\varphi) ,g^{\!_{\Psi}}(\varphi) \}.
\label{optimalFprobabilistic}
\end{equation}
By ranging $\varphi$ from $0$ to  $\pi$ we can obtain the optimal settings for all instances listed in 
Eqs.~(\ref{1of8})-(\ref{8of8}), and by choosing the greatest value from $g^{\!_{\Phi}}(\varphi_{opt})$ and
$g^{\!_{\Psi}}(\tilde{\varphi}_{opt})$, we get the optimal efficiency $ \overline{F}_{\!_{opt}}$. The corresponding success
rate is given by either $2 q(\varphi_{opt})$ or $2 q(\tilde{\varphi}_{opt})$, where
$\varphi_{opt}$ and $\tilde{\varphi}_{opt}$ are the $\varphi$'s maximizing $g^{\!_{\Phi}}$
and $g^{\!_{\Psi}}$, respectively.

\section{Results}
\label{results}

We are now ready to study the efficiency to teleport an arbitrary pure state qubit 
for several entangled resources described by Heisenberg-like models 
in thermal equilibrium with a heat reservoir at temperature $T$.  
We divide our entangled resources into two main groups, all of which 
subjected to external magnetic fields in the $z$ direction. The first group encompasses
all models in which there is no $\sigma_z^{(2)}\sigma_z^{(3)}$ interaction and the second one
those models possessing it. Note that whenever there is no external fields, the deterministic and
probabilistic protocols yield the same results and, thus, we work only with cases in which the 
external field is present. 

\subsection{XY-like models}

The one-dimensional XY model in a transverse field is obtained from Eq.~(\ref{hamiltonian}) 
by setting $j_z=0$ and $h_a=h_b$. It is more usual, however, to rewrite Eq.~(\ref{hamiltonian})
as follows \cite{werIJMPB},
\begin{equation}
H = -\lambda[(1+\zeta)\sigma_x^{(2)} \sigma_x^{(3)}+(1-\zeta)\sigma_y^{(2)} \sigma_y^{(3)}]
-\sigma_z^{(2)} - \sigma_z^{(3)},
\label{hamiltonian2} 
\end{equation}
with $\lambda\geq 0$ being the inverse of the magnitude of the external field and $\zeta$ the anisotropy
parameter. The Ising model is obtained when $\zeta = \pm 1$ and for $\zeta = 0$ we get the XX model
in a transverse field. At $T=0$ and in the thermodynamic limit (infinite chain), 
the XY model has a quantum critical point at $\lambda = 1$, 
where a second-order quantum phase transition separates a 
ferromagnetic ordered phase from a paramagnetic one \cite{lie61,bar70,bar71,pfe70}.

\subsubsection{Efficiency as a function of $T$}

\begin{figure}[!ht]
\includegraphics[width=8.5cm]{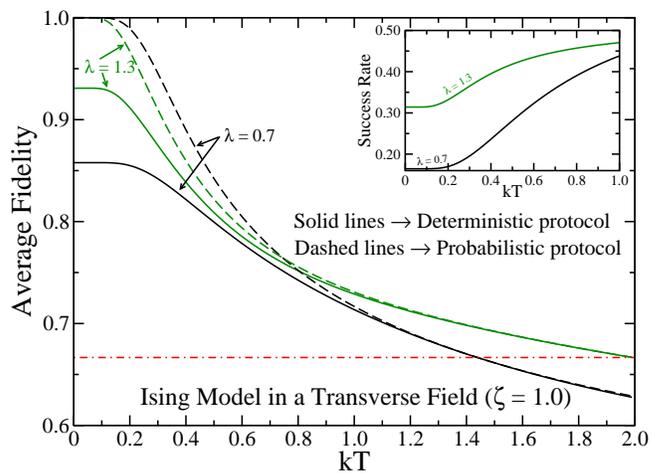}
\caption{\label{figIsing}(color online) Main plot: The efficiencies for the 
deterministic (solid curves) and probabilistic (dashed curves) teleportation protocols as a function of the temperature
when the quantum communication channel connecting Alice and Bob is given by the thermalized Ising model in a transverse field. 
The efficiency for the deterministic protocol is given by Eq.~(\ref{optimalFdeterministic}) and for the probabilistic
one by Eq.~(\ref{optimalFprobabilistic}).
The dotted-dashed red line marks the classical limit ($2/3$) below which the 
teleportation protocol can be matched by a purely classical protocol. 
Inset: The success rate (probability of success) for the probabilistic protocol. 
Here and in the following figures all quantities are dimensionless.}
\end{figure}

We first analyze the optimal efficiencies (average fidelities) of the deterministic and probabilistic
protocols, Eqs.~(\ref{optimalFdeterministic}) and (\ref{optimalFprobabilistic})
respectively, as a function of the temperature $T$. We start with the Ising model in a transverse field,
whose main results are shown in Fig.~\ref{figIsing}.

Looking at Fig.~\ref{figIsing} we note that the efficiency is a monotonically decreasing function  
of the temperature and that for $kT \approx 1.2$ the optimal efficiencies for
the deterministic and probabilistic protocols are almost the same. As we continue to increase the temperature,
we arrive at a value of $T$ after which the efficiency of the protocol is below $2/3$. This value for the average
fidelity is called the classical limit since any protocol with average fidelities lower than $2/3$ can be implemented
without Alice and Bob sharing an entangled resource \cite{mas95}. See also the Appendix for further details and
a proof of this limit for the deterministic teleportation protocol studied here.

For low values of $T$,
however, we can have considerable gains in efficiency by working with the probabilistic protocol.
For instance, whenever $kT < 0.2$, the probabilistic protocol yields an almost perfect teleportation,
a considerable improvement over the deterministic one. In this case the success rate 
is about $10\%$ when $\lambda = 0.7$ and $30\%$ when $\lambda = 1.3$. We also note that the optimal 
efficiencies for the deterministic and probabilistic protocols are given by $f^{\!_{\Phi}}_{opt}$ and
$g^{\!_{\Phi}}(\varphi)$, respectively (see Eqs.~(\ref{optimalFdeterministic}) and (\ref{optimalFprobabilistic})).
Moreover, the optimal $\varphi$ for the latter depends on $T$ and is not equal to $\pm \pi/4$. 

\begin{figure}[!ht]
\includegraphics[width=8.5cm]{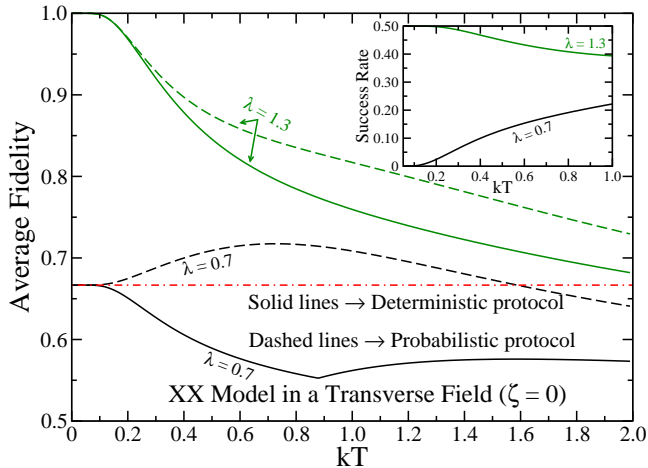}
\caption{\label{figXX}(color online) Same as Fig.~\ref{figIsing} but now we work with 
the isotropic XX model in a transverse field. Note that under certain conditions ($\lambda < 1$) the
efficiency for the probabilistic protocol may increase with the temperature 
and be the only one yielding an efficiency greater than the classical threshold (2/3). Also,
the optimal $\varphi$ for the probabilistic protocol depends on $T$ and is not equal to $\pm \pi/4$,
with the latter being the optimal settings for the deterministic case.}
\end{figure}

\begin{figure}[!ht]
\includegraphics[width=8.5cm]{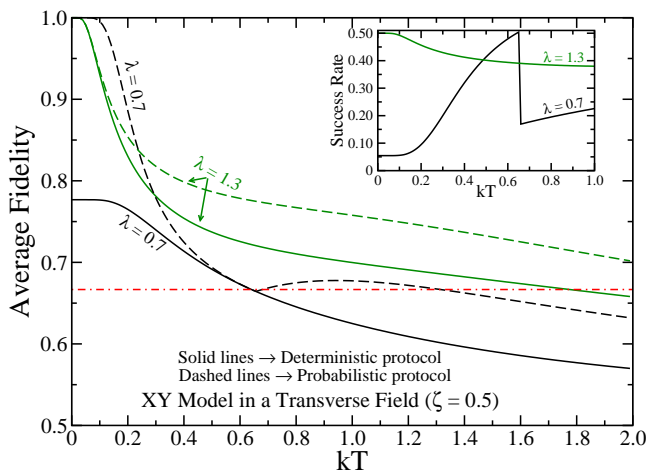}
\caption{\label{figXY}(color online) Same as Figs.~\ref{figIsing} and \ref{figXX} but now we have 
the anisotropic XY model in a transverse field.}
\end{figure}

Moving to the XX and XY models, i.e., turning on the $\sigma_y^{(2)} \sigma_y^{(3)}$ interaction,
we observe the following two similar and interesting trends (see Figs.~\ref{figXX} and \ref{figXY}).
First, whenever $\lambda < 1$ (ferromagnetic phase) there exists a range of values of 
temperature where the efficiency of the probabilistic protocol \textit{increases} with $T$.
This is a remarkable property and tells us that working with a ``warmer'' entangled resource is better than
working with a ``colder'' one. We can understand this behavior noting that under certain configurations 
of the coupling constants, the ground state of the Hamiltonian has little or no entanglement at all, although
the first excited states are highly entangled ones and very close to Bell states \cite{Rig03}. 
Thus, by increasing the temperature we start to populate those highly entangled states in such a manner that 
a warmer entangled resource has more entanglement than a colder one. The latter effect is more intense in the probabilistic
protocol where, by postselecting the appropriate measurement results, we may project the entangled resource $\rho_{ch}$ onto 
highly entangled states and consequently enhance even more the efficiency of the teleportation protocol. 
If we continue to increase the temperature, however,
more and more states get populated and we start to get a less entangled quantum communication channel, reducing the efficiency of the protocol.
For sufficiently high temperatures
the entangled resource is nearly described by a completely mixed state with no entanglement at all. This is why we always
end up with efficiencies lower than $2/3$ for very high temperatures.

Second, another important characteristic shared by the XX and XY models is the
fact that for certain values of $T$ the efficiency for the deterministic protocol 
does not surpass the classical limit $2/3$, while the probabilistic protocol's efficiency does. 
In this scenario, therefore, we can only get a truly quantum teleportation if we employ 
the probabilistic protocol.  

There are also different characteristics between the XX and XY models. For example,
the deterministic protocol for the XX model does not yield an average fidelity greater
than the classical limit for $\lambda <1$. This is only possible when we use the probabilistic protocol. 
For the XY model, however, there is no such restriction and  
we can have for $\lambda < 1$ the average fidelity for both the deterministic and probabilistic protocols greater 
than $2/3$ if we work at a sufficiently low temperature. 

Another distinctive feature of the XX model is the fact that whenever the optimal average fidelities for 
the deterministic and probabilistic protocols are greater than $2/3$, 
$f^{\!_{\Psi}}_{opt}$ and $g^{\!_{\Psi}}(\varphi)$, respectively, are the functions optimizing 
the efficiency (see Eqs.~(\ref{optimalFdeterministic}) and (\ref{optimalFprobabilistic})).
For the XY model, however, the functions leading to the optimal efficiency for certain values of $T$ may be different 
for the probabilistic protocol when the efficiency is greater than $2/3$.  In this case either $g^{\!_{\Phi}}(\varphi)$ or 
$g^{\!_{\Psi}}(\varphi)$ may give the optimal efficiency. This is the reason for the 
cusp of the curve of the optimal average fidelity (the $\lambda = 0.7$ dashed curve)
and for the discontinuity in the success rate (the $\lambda = 0.7$ curve in the inset) that we
see for the probabilistic protocol in Fig.~\ref{figXY}. The cusp for the efficiency curve and
the discontinuity for the probability of success curve occur exactly at the temperature 
in which $g^{\!_{\Phi}}(\varphi)$ and $g^{\!_{\Psi}}(\varphi)$ exchange roles. Below this
temperature $g^{\!_{\Phi}}(\varphi)$ gives the optimal efficiency while above it  
$g^{\!_{\Psi}}(\varphi)$ does.

\subsubsection{Efficiency as a function of the external field}

\begin{figure}[!ht]
\includegraphics[width=8.5cm]{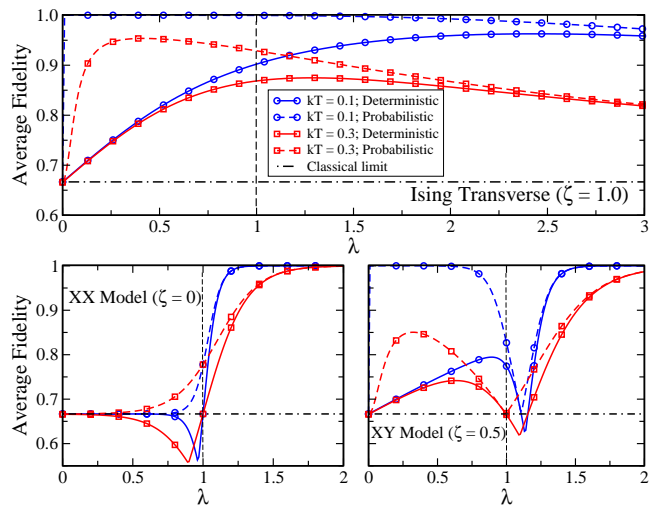}
\caption{\label{figXYlambda}(color online) The optimal efficiencies of the deterministic (solid curves)
and probabilistic (dashed curves) teleportation protocols as a function of $\lambda$ 
for the Ising model (upper panel), the XX model (left lower panel),
and the XY model (right lower panel), all of them in an external magnetic field of strength $1/\lambda$.
Circles denote $kT = 0.1$ and squares $kT = 0.3$. The dotted-dashed black lines delimit the classical
limit $2/3$.}
\end{figure}

We now turn our attention to the behavior of the average fidelities for the deterministic
and probabilistic protocols as functions of the inverse of the strength of the external
magnetic field $\lambda$. Starting with the Ising model in a transverse field (upper panel
of Fig.~\ref{figXYlambda}), we note that for a fixed temperature there is an optimal $\lambda$
that gives the greatest efficiency and that the optimal $\lambda$'s are different for the 
deterministic and probabilistic protocols. This is most clearly seen looking at the curves 
for $kT = 0.3$. We also see that the probabilistic protocol outperforms by far the deterministic
one for small values of $\lambda$.

Studying the XX model (left lower panel of Fig.~\ref{figXYlambda}), we note that the greater the 
value of $\lambda$ the better the efficiency of the probabilistic protocol. For the deterministic
protocol, an increase of $\lambda$ increases the efficiency only for $\lambda$ greater than a 
certain critical value that depends on $T$. Also, when
$\lambda < 1$ the average fidelity for the deterministic protocol does not exceed $2/3$. 
It is interesting to note that the efficiencies for the probabilistic protocols, and in particular 
for the deterministic ones, change
abruptly near the quantum critical point $\lambda = 1$. 

Near the quantum critical point $\lambda = 1$ there is a similar abrupt behavior for the 
efficiencies of the deterministic and probabilistic 
protocols for the XY model (right lower panel of Fig.~\ref{figXYlambda}). 
In this case the average fidelities tend to their 
minimum values near the critical point. As we move to the right or left of the critical 
point the efficiency starts to increase. For $\lambda > 1$ this trend continues as we 
increase $\lambda$ while for $\lambda < 1$ the average fidelity starts to decrease after reaching
a local maximum. This behavior is clearer the greater the value of $T$. Finally, the reason for the
cusps in the curves for the efficiencies is again related to which of the functions
$f^{\!_{\Phi}}_{opt}$ or $f^{\!_{\Psi}}_{opt}$ ($g^{\!_{\Phi}}(\varphi)$ or 
$g^{\!_{\Psi}}(\varphi)$) gives the optimal average fidelity for the deterministic 
(probabilistic) protocol. For small $\lambda$, $f^{\!_{\Phi}}_{opt}$ and
$g^{\!_{\Phi}}(\varphi)$ give the highest efficiencies and, as we increase $\lambda$, $f^{\!_{\Psi}}_{opt}$ and
$g^{\!_{\Psi}}(\varphi)$ dominate after we cross a certain value of $\lambda$ that depends
on $T$.

\subsection{XXZ-like models}

The one-dimensional XXZ model in an external field in the $z$ direction is obtained from
Eq.~(\ref{hamiltonian}) when we set $j_x=j_y$ and $h_a=h_b$. This model is usually written as
\cite{werIJMPB}
\begin{equation}
H = 2J[\sigma_x^{(2)} \sigma_x^{(3)}+\sigma_y^{(2)} \sigma_y^{(3)}+\Delta \sigma_z^{(2)} \sigma_z^{(3)}]
-\frac{h}{2} [\sigma_z^{(2)}+\sigma_z^{(3)}],
\label{hamiltonian3}
\end{equation}
where $J$ is the exchange constant, $\Delta$ the anisotropy parameter, and $h$ the external field.
When $\Delta = 1$ we have the isotropic XXX model and for $\Delta \neq 0$ we get the anisotropic XXZ
model. In the thermodynamic limit and at $T=0$ the XXZ model has two quantum critical points \cite{clo66,tak99,bor05,boo08,tri10}:
$\Delta_{inf}$, where an infinite order quantum phase transition takes place, and $\Delta_1$, where a first-order
quantum phase transition happens. The expressions giving those critical points are not so simple and can be found
in Refs. \cite{clo66,tak99}. 

\subsubsection{Efficiency as a function of $T$}

\begin{figure}[!ht]
\includegraphics[width=8.5cm]{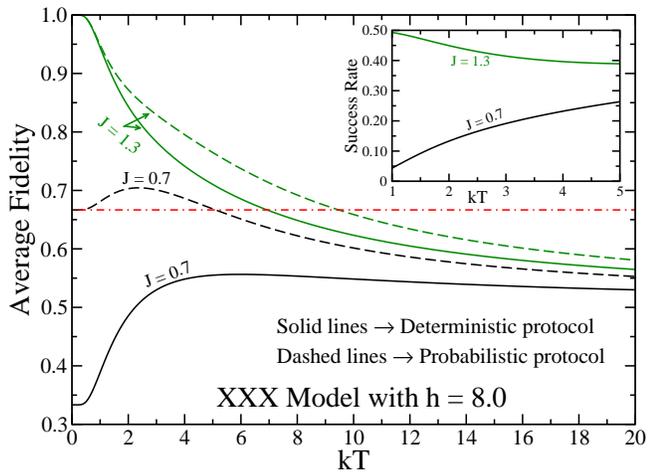}
\caption{\label{figXXX}(color online) Main plot: The average fidelities (efficiencies) of the 
deterministic (solid curves) and probabilistic (dashed curves) teleportation protocols as a function of the temperature
when the quantum communication channel connecting Alice and Bob is the thermalized XXX model in an external field. 
As explained in the text, for the deterministic protocol we plot $f^{\!_{\Psi}}_{opt}$ and for the probabilistic
one Eq.~(\ref{optimalFprobabilistic}).
The dotted-dashed red line marks the classical limit ($2/3$) below which the 
teleportation protocol can be matched by a purely classical protocol. 
Inset: The success rate (probability of success) for the probabilistic protocol.}
\end{figure}

Let us start studying the isotropic XXX model ($\Delta = 1$). The first thing worth noting 
is that for $J<0$ the efficiencies for both the deterministic and probabilistic protocols do not
surpass the classical limit $2/3$, even for low $T$. We thus restrict the following analysis to the cases
in which $J>0$. It can also be proved that for the deterministic protocol $f^{\!_{\Phi}}_{opt}\leq 2/3$ and thus,
since we are interested in the cases surpassing the classical limit, instead of Eq.~(\ref{optimalFdeterministic})
we work only with $f^{\!_{\Psi}}_{opt}$ in the determination of the optimal efficiency.
The curves for the deterministic protocol in Figs.~\ref{figXXX} and \ref{figXXZDel} show $f^{\!_{\Psi}}_{opt}$.
Also, by setting the magnetic field to $h=8.0$ we get that the first-order quantum phase transition for this
model occurs at $J_c=1.0$.

Looking at Fig.~\ref{figXXX} we note that we have two regimes 
for the behavior of the average fidelities. For $J < J_c$ 
the deterministic protocol does not give an efficiency greater than $2/3$. 
In this regime the classical limit can only be surpassed using the 
probabilistic protocol. Indeed, for $kT \lesssim 5.0$ the probabilistic protocol yields
an efficiency greater than $2/3$ and greater than that of the deterministic protocol,
with success rates of the order of $10\%$.
We also see that in this range of temperatures
there are instances where the efficiency \textit{increases} with $T$. 

For $J>J_c$, on the other hand, both the deterministic and probabilistic
protocols can yield efficiencies above the classical limit. In this regime
the efficiencies are always a monotonically decreasing function of the temperature and 
we still have a small range of temperatures in which only the probabilistic protocol
gives an average fidelity greater than $2/3$. 
For all values of $J>0$ the optimal efficiency for the probabilistic protocol is given by $g^{\!_{\Psi}}(\varphi)$, 
with the optimal $\varphi$ being different from $\pm \pi/4$ and dependent on $T$.

\begin{figure}[!ht]
\includegraphics[width=8.5cm]{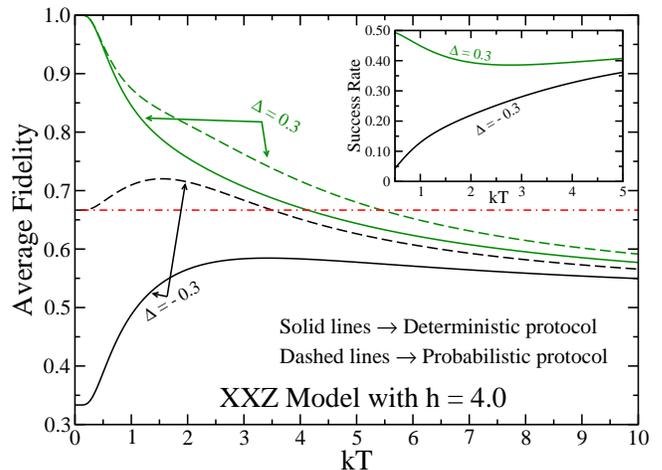}
\caption{\label{figXXZ}(color online) Same as Fig.~\ref{figXXX} but now we work with
the XXZ model with an external field in the $z$ direction.}
\end{figure}

We now focus our attention at the XXZ model in an external field in the $z$ direction.
We set $J=1.0$ and the magnitude of the field ($h=4.0$) such that
the first-order quantum phase transition occurs at $\Delta_1 = 0$. Here we can also
prove that $f^{\!_{\Phi}}_{opt}\leq 2/3$ for the deterministic protocol and similarly to
the XXX model, we show $f^{\!_{\Psi}}_{opt}$ instead of Eq.~(\ref{optimalFdeterministic})
in Figs.~\ref{figXXZ} and \ref{figXXZDel} when analyzing the deterministic protocol.

Looking at Fig.~\ref{figXXZ} we note that many features seen for the XXX model 
are also present in the XXZ model. Indeed, we have two regimes for the behavior
of the efficiency of the protocol. One before ($\Delta < \Delta_1$) and another
after ($\Delta > \Delta_1$) the quantum critical point delimiting the first-order
quantum phase transition. For $\Delta < \Delta_1$ only the probabilistic protocol
yields average fidelities greater than the classical limit, with success rates
lying between $10\%$ to $20\%$ for a considerable set of values of $\Delta < \Delta_1$.
We also have ranges of temperatures where the efficiency of the probabilistic protocol 
\textit{increases} with $T$. 

For $\Delta > \Delta_1$ the efficiencies are monotonically decreasing functions of $T$
and both the deterministic and probabilistic protocols can work above the classical limit
at sufficiently low temperatures. We also have small ranges of $T$
in which the probabilistic protocol leads to an efficiency greater than $2/3$ while the
deterministic protocol works below this value. 

Finally, and similarly to the XXX model, whenever the efficiency is above the classical limit,
the functions leading to the optimal efficiencies
are $f^{\!_{\Psi}}_{opt}$ for the deterministic and $g^{\!_{\Psi}}(\varphi)$ for the probabilistic protocols. 
The optimal values of $\varphi$ for the probabilistic protocol depend on $T$ and are not $\pm \pi/4$, the optimal ones for
the deterministic case.

\subsubsection{Efficiency as a function of the coupling constants}

We now investigate how the efficiencies (average fidelities) for the deterministic
and probabilistic protocols behave as a function of the exchange constant $J$ for the XXX model and
of the anisotropy parameter $\Delta$ for the XXZ model.

\begin{figure}[!ht]
\includegraphics[width=8.5cm]{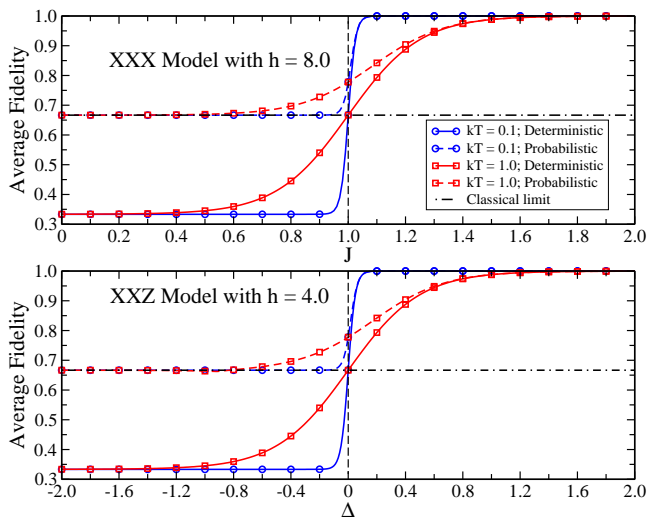}
\caption{\label{figXXZDel}(color online) The optimal efficiencies of the deterministic (solid curves)
and probabilistic (dashed curves) teleportation protocols as a function of $J$ 
for the XXX model (upper panel) and of $\Delta$ for the XXZ model (lower panel).
As discussed in the text, for the deterministic protocol we plot $f^{\!_{\Psi}}_{opt}$ and for the probabilistic
one Eq.~(\ref{optimalFprobabilistic}).
Circles denote $kT = 0.1$ and squares $kT = 1.0$. The dotted-dashed black lines delimit the classical
limit $2/3$.}
\end{figure}

For the XXX model we keep as before $h=8.0$ and for several values of $kT$ we compute the 
efficiency as a function of $J$ (upper panel of Fig.~\ref{figXXZDel}), including values of $J$
near and at the critical point $J_c=1.0$. 
For the XXZ model we set $h=4.0$ and $J=1.0$, which leads to a critical point $\Delta_1=0$, and we also evaluate
for several values of $kT$ the efficiency as a function of $\Delta$ (lower panel of Fig.~\ref{figXXZDel}), 
including values of $\Delta$ near and at the critical point $\Delta_1$.

Looking at Fig.~\ref{figXXZDel} we note that the efficiencies for the XXX and XXZ models, as functions of 
$J$ and $\Delta$, respectively, share the same qualitative features. 
In particular, we note a clear distinctive behavior for the optimal average fidelities before and after
the first-order quantum critical points, even at considerably high temperatures ($kT \approx 1.0$).
It is now clear that below the critical point the deterministic protocols do not yield an efficiency
greater than the classical limit $2/3$ while the probabilistic protocols do. 
We also observe that for sufficiently high values of $J$ and $\Delta$, above the critical points,
the efficiencies for the deterministic and probabilistic protocols converge to their greatest possible 
value, leading to a perfect teleportation. 
Moreover, at low values of $J$ or $\Delta$, the functions $f^{\!_{\Phi}}_{opt}$ and $g^{\!_{\Phi}}(\varphi)$ 
give the optimal average fidelities. As we approach the critical point, $f^{\!_{\Psi}}_{opt}$ and
$g^{\!_{\Psi}}(\varphi)$ dominate and furnish the optimal values for the efficiencies. Note that this exchange of 
functions leading to the optimal efficiencies never occurs exactly at the critical point for finite $T$.

It is also worth noting that we have computed the efficiencies
about and at the other quantum critical point, where an infinite-order quantum phase transition happens ($\Delta_{inf}$).
For the present XXZ model with $h=4.0$ and $J=1.0$ we obtain $\Delta_{inf}\approx 2.74$ \cite{clo66,tak99,werPRA}. 
We have not observed, however,
any quantitative or qualitative changes in the behavior of the efficiencies. Actually, before reaching 
$\Delta_{inf}$ the efficiency of the teleportation protocol already saturates to its highest possible value and
no changes are seen after that value is attained.

\section{Conclusion}

We have extensively studied the probabilistic teleportation protocol when the
entangled resource connecting Alice and Bob is given by interacting two-qubit systems in equilibrium with a thermal reservoir.
In this scenario the quantum state describing the entangled resource is the canonical ensemble density matrix and any
entanglement present in that state is usually dubbed 
\textit{thermal entanglement} \cite{Ved01,Kam02,Rig03,Ami07,reviews,Wer10,werPRL,werPRA,werIJMPB}.

We worked with several standard Heisenberg-like models in order to describe the interaction between 
the two qubits of the quantum communication channel. 
Those models are widely employed to describe the interaction between
two or more spins in several condensed matter systems and can be used to describe the interactions we
might face when building a quantum computer or a quantum communication protocol operating 
on solid state devices.  
Being more specific, we studied the Ising model, 
the XX model, the XY model, the isotropic XXX model, and the anisotropic XXZ model. 
We also considered the cases where an external magnetic field is applied in the $z$ direction.  

After studying all those models three important common features emerged. First, we proved
analytically that the efficiency for the probabilistic protocol can only be greater than the 
efficiency of the deterministic protocol if we have an external magnetic field. Whenever 
the external field is zero, the probabilistic and deterministic protocols have exactly 
the same efficiency. 

Second, whenever the probabilistic teleportation protocol outperforms 
the deterministic protocol, the measurement basis employed by Alice during the execution of
the teleportation protocol is not the standard Bell basis, which is spanned by four maximally entangled states.
The optimal measurement basis for the probabilistic protocol is given by the generalized Bell states, 
whose entanglement degree is not maximal. Moreover, the appropriate generalized Bell basis
depends on the value of the temperature and on which Heisenberg-like model we are working with. 

Third, the optimal settings leading to the optimal efficiency
for the probabilistic protocol are the same for two out of four possible measurement results that
Alice may obtain at each run of the protocol. 
Thus, the success rate for the probabilistic protocol is enhanced since Alice and Bob 
can postselect two instead of one measurement result at each run of the protocol. In general,
the success rate for the probabilistic protocols here studied are above $10\%$, being much higher
than this value under certain arrangements.

Other three features are clearly shared by all models here investigated with the
exception of the Ising model. The first one is related to the fact that \textit{more} heat 
(higher temperatures) may lead to a \textit{more} efficient probabilistic teleportation. In the notation of the
present paper, this happens 
whenever the coupling constants and the external magnetic field are such that the system lies below
the quantum critical point separating its two phases. In the appropriate phase,
there exists a scenario in which the efficiency increases with increasing temperature. 

Another characteristic shared by almost all models is the fact that under the same conditions the 
optimal efficiencies for the probabilistic and deterministic protocols may differ in a very important way. 
There are ranges of temperatures where only the probabilistic protocol crosses the classical limit of $2/3$ 
for the optimal average fidelity. Below this value any teleportation protocol can be simulated by a 
``classical'' protocol, where no entanglement at all is needed between Alice and Bob. 
Only local operations and classical communication (LOCC) suffice to deliver the same efficiency. 
Thus, whenever this happens, we can only have a truly quantum teleportation if we work with 
the probabilistic protocol. The deterministic protocol fails in delivering a 
quantum teleportation that is genuinely quantum.

Third, we have also noted that the behavior for the efficiencies of the deterministic and probabilistic
protocols may be qualitatively and quantitatively affected in the vicinity of 
the quantum critical points, even at finite temperatures. 
For instance, for the XX, XXX and XXZ models the optimal efficiencies can only surpass the 
classical limit $2/3$ as we approach the critical point from below. The lower the temperature
the more the quantum critical point marks this transition in the behavior for the efficiency.
For the XX and XY models, we also noted that near the critical point we have the global minimum for 
the efficiency.

Finally, and similarly to the results of Ref. \cite{for16}, we have a trade-off between the 
success rates and the efficiencies for the probabilistic protocols. 
The optimizations performed here were carried out to maximize the average fidelity without imposing any other restriction.
It is possible, however, to increase the success rate by diminishing the efficiency.

\begin{acknowledgments}
RF thanks CAPES (Brazilian Agency for the Improvement of Personnel of Higher Education)
for funding and GR thanks the Brazilian agencies CNPq
(National Council for Scientific and Technological Development) and
CNPq/FAPESP (State of S\~ao Paulo Research Foundation) for financial support through the National Institute of
Science and Technology for Quantum Information.
\end{acknowledgments}

\appendix*

\section{The classical limit for the average fidelity}

Our goal here is to prove that the average fidelity as
given by Eq.~(\ref{deterministicF}) for the deterministic teleportation protocol cannot have
values greater than $2/3$ if Alice and Bob share a non-entangled state. 

The most general non-entangled state that Alice and Bob can share is given by \cite{wer89}
\begin{equation}
\rho^{AB} = \sum_{k=1}^n p_k \rho^A_k \otimes \rho^B_k,
\label{noEntanglement}
\end{equation}
where $n$ is a positive integer, $0\leq p_k \leq 1$, $\sum_{k=1}^n p_k = 1$,
and $\rho^A_k$ and $\rho^{B}_k$ are density matrices describing states with Alice and 
Bob, respectively. Equation (\ref{noEntanglement}) is a convex combination of 
product states.

Due to the linearity of the two averaging processes employed to define Eq.~(\ref{deterministicF}) 
and to the fact that Eq.~(\ref{noEntanglement}) is a convex combination of product states we obtain
\begin{equation}
\langle \overline{F} \rangle_{\!_{\rho^{AB}}} = \sum_{k=1}^{n} p_k \langle \overline{F} \rangle_{\!_{\rho^{A}_k\otimes \rho^{B}_k}}.
\label{average1}
\end{equation}
The subscripts $\rho^{AB}$ and $\rho^{A}_k\otimes \rho^{B}_k$ attached to $\langle \overline{F} \rangle$
tell us which shared quantum resource between Alice and Bob we are employing to compute the averages. We should
also note that a long but straightforward calculation, where we use Eq.~(\ref{noEntanglement}) to compute 
Eqs.~(\ref{prob}), (\ref{twelve}), (\ref{Fj}), and finally (\ref{deterministicF}), also 
leads to Eq.~(\ref{average1}).

As we will show in what follows,
\begin{equation}
\langle \overline{F} \rangle_{\!_{\rho^{A}_k\otimes \rho^{B}_k}} \leq 2/3.
\label{inequality1}
\end{equation}
Thus, inserting Eq.~(\ref{inequality1}) into (\ref{average1}) we get
\begin{equation}
\langle \overline{F} \rangle_{\!_{\rho^{AB}}} \leq \frac{2}{3}\sum_{k=1}^{n} p_k = \frac{2}{3},
\end{equation}
which proves our claim.

In order to prove Eq.~(\ref{inequality1}) we first note that the most general way
of writing a density matrix describing a single qubit is
\begin{equation}
\rho^A = (\mathbb{1}+a_x\sigma_x^A+a_y\sigma_y^A+a_z\sigma_z^A)/2.
\label{matrixA}
\end{equation}
Here $A$ denotes Alice, $a_j=\mbox{Tr}[\sigma_j^A\rho^A]$ for $j=x,y,z$, 
the symbol $\mathbb{1}$ is the unitary matrix of dimension $2$, and $\sigma_j^A$
are the Pauli matrices. A similar expression can be written for Bob,
\begin{equation}
\rho^B = (\mathbb{1}+b_x\sigma_x^B+b_y\sigma_y^B+b_z\sigma_z^B)/2.
\label{matrixB}
\end{equation}

The eigenvalues of $\rho^A$ are 
$$
\lambda_{\pm} = \left(1\pm \sqrt{a_x^2+a_y^2+a_z^2}\right)/2.
$$
Since $\rho^A$ is positive definite and normalized to one we must have $0\leq\lambda_{\pm}\leq 1$,
which implies that
\begin{equation}
a_x^2+a_y^2+a_z^2 \leq 1.
\label{inequalityA}
\end{equation}
A similar argument for $\rho^B$ gives
\begin{equation}
b_x^2+b_y^2+b_z^2 \leq 1.
\label{inequalityB}
\end{equation}

Now, if we use Eqs.~(\ref{matrixA}) and (\ref{matrixB}) to compute $\rho^A\otimes \rho^B$ and use it 
in the evaluation of Eq.~(\ref{deterministicF}) we get
\begin{eqnarray}
\!\!\!\!\!\langle \overline{F} \rangle_{\!_{\rho^{A}\otimes \rho^{B}}}^{\!^{\Phi^{\pm}}} = [3 + a_zb_z\pm(a_xb_x-a_yb_y)\sin(2\varphi)]/6, \\
\!\!\!\!\!\langle \overline{F} \rangle_{\!_{\rho^{A}\otimes \rho^{B}}}^{\!^{\Psi^{\pm}}} = [3 - a_zb_z\pm(a_xb_x+a_yb_y)\sin(2\varphi)]/6,
\end{eqnarray}
where the superscripts, as explained in Sec. \ref{heisenberg}, denote the four possible sets
of unitary corrections that Bob can apply to his qubit. The parameter $\varphi$ defines which 
generalized Bell states Alice uses to project her qubits (see Sec. \ref{densitymatrix}).

Since $\varphi$ can be freely set by Alice, she can always choose it to maximize the above expressions, leading
to the following optimal average fidelities for the deterministic teleportation protocol,
\begin{eqnarray}
\langle \overline{F} \rangle_{\!_{\rho^{A}\otimes \rho^{B},opt}}^{\!^{\Phi^{\pm}}} = (3 + a_zb_z+|a_xb_x-a_yb_y|)/6, \\
\langle \overline{F} \rangle_{\!_{\rho^{A}\otimes \rho^{B},opt}}^{\!^{\Psi^{\pm}}} = (3 - a_zb_z+|a_xb_x+a_yb_y|)/6,
\end{eqnarray}
where $|x|$ is the magnitude of $x$.
Those optimal average fidelities satisfy the following inequality, which is an upper bound for their possible 
values ($\epsilon = \Phi^{\pm},\Psi^{\pm}$),
\begin{equation}
\langle \overline{F} \rangle_{\!_{\rho^{A}\otimes \rho^{B},opt}}^{\!^{\epsilon}} \leq  (3 + |a_xb_x| + |a_yb_y| + |a_zb_z|)/6.
\end{equation}
But as we show below, 
\begin{equation}
|a_xb_x| + |a_yb_y| + |a_zb_z| \leq 1,
\label{inequality2}
\end{equation}
leading to the proof of Eq.~(\ref{inequality1}),
\begin{equation}
\langle \overline{F} \rangle_{\!_{\rho^{A}_k\otimes \rho^{B}_k}} \leq 
\langle \overline{F} \rangle_{\!_{\rho^{A}\otimes \rho^{B},opt}}^{\!^{\epsilon}} \leq  (3 + 1)/6 = 2/3.
\end{equation}

We can show that Eq.~(\ref{inequality2}) is indeed true by noting that the sum of the following three identities,
\begin{eqnarray}
(|a_x|-|b_x|)^2 \geq 0 &\Longrightarrow&  |a_xb_x| \leq (a_x^2+b_x^2)/2, \\
(|a_y|-|b_y|)^2 \geq 0 &\Longrightarrow&  |a_yb_y| \leq (a_y^2+b_y^2)/2, \\
(|a_z|-|b_z|)^2 \geq 0 &\Longrightarrow&  |a_zb_z| \leq (a_z^2+b_z^2)/2,
\end{eqnarray}
gives
\begin{equation}
|a_xb_x| + |a_yb_y| + |a_yb_y| \leq (a_x^2+a_y^2+a_z^2+b_x^2+b_y^2+b_z^2)/2.
\end{equation}
Then, using Eqs.~(\ref{inequalityA}) and (\ref{inequalityB}) we immediately get
\begin{equation}
|a_xb_x| + |a_yb_y| + |a_yb_y| \leq (1+1)/2 = 1,
\end{equation}
which proves Eq.~(\ref{inequality2}).

\textit{Remarks.} It is important to note that the previous proof can be extended to
arbitrary projective measurements that Alice might implement onto her qubits and
also to arbitrary sets $\{ V_1,V_2,V_3,V_4\}$ of unitary operations that Bob might apply
to his qubits, as long as $V_j$ are orthogonal in the sense that the Hilbert-Schmidt inner product
between different matrices are zero. More specifically, we must have $\mbox{Tr}[V_jV_k^\dagger] = 2\delta_{jk}$.
Note that the original set of the standard teleportation protocol, for example,
$\{ U_1,U_2,U_3,U_4\}=\{ \mathbb{1},\sigma_z,\sigma_x,\sigma_z\sigma_x\}$ are orthogonal in the above sense and,
as we will see, that is why $V_j$ inherits this property.

Let us start showing that the previous proof applies to arbitrary projective measurements. First we write
Eq.~(\ref{noEntanglement}) as follows,
\begin{equation}
\tilde{\rho}^{AB} = \sum_{k=1}^n p_k \tilde{\rho}^A_k \otimes \rho^B_k,
\label{noEntanglement2}
\end{equation}
where $\tilde{\rho}^A_k=U^A\rho^A_k (U^A)^\dagger$ and $U^A$ is an 
arbitrary unitary operator acting on the Hilbert space of qubit $A$. 
We also write the input qubit to be teleported as 
$\tilde{\rho}_{in} =U_{in}\rho_{in}U_{in}^\dagger$, with $U_{in}$ an
arbitrary unitary operator acting on the Hilbert space of the input qubit.
With those choices, the total state describing the input qubit and $\tilde{\rho}^{AB}$ is
\begin{equation}
\tilde{\rho} = (U_{in} \otimes U^A) (\rho_{in} \otimes \rho^{AB} ) (U_{in}^\dagger \otimes (U^A)^\dagger). 
\label{rhotilde}
\end{equation}

The state $\tilde{\rho}$ changes to 
\begin{equation}
\tilde{\rho} \longrightarrow \frac{P_j^\varphi \tilde{\rho} P_j^\varphi}{\mbox{Tr}[P_j^\varphi \tilde{\rho}P_j^\varphi]} 
\end{equation}
after Alice projects her state onto the generalized Bell state $|B_j^\varphi\rangle$, Eqs.~(\ref{B1})-(\ref{B4}),
with projector $P_j^\varphi$ given by Eq.~(\ref{Bj}). Tracing out Alice's qubits we get the state with Bob (before he applies 
his unitary correction),
\begin{equation}
\tilde{\rho}_{\!_{B_j}} = \frac{\mbox{Tr}_{in,A}[P_j^\varphi \tilde{\rho} P_j^\varphi]}{\mbox{Tr}[P_j^\varphi \tilde{\rho}P_j^\varphi]}
= \frac{\mbox{Tr}_{in,A}[P_j^\varphi \tilde{\rho}]}{\mbox{Tr}[P_j^\varphi \tilde{\rho}]},
\label{bobj}
\end{equation}
where the last equation was obtained using the invariance of the trace under cyclic permutations and that
$P_j^\varphi P_j^\varphi = P_j^\varphi$. Inserting Eq.~(\ref{rhotilde}) into (\ref{bobj}) and once again 
using the invariance of the trace under cyclic permutations we get
\begin{equation}
\tilde{\rho}_{\!_{B_j}} = \frac{\mbox{Tr}_{in,A}[\tilde{P}_j\rho ]}
{\mbox{Tr}[\tilde{P}_j\rho]},
\label{bobj2}
\end{equation}
where $\rho = \rho_{in}\otimes \rho^{AB}$ and 
\begin{equation}
\tilde{P}_j = (U_{in}^\dagger \otimes (U^A)^\dagger)P_j^\varphi (U_{in} \otimes U^A).
\label{Pjtilde}
\end{equation}
Now, if we show that $\tilde{P}_j$ can represent an arbitrary projector, 
we have shown that the proof of the classical limit is valid for arbitrary projective measurements.

The key tool we need to show that $\tilde{P}_j$ is an arbitrary projector is the Schmidt decomposition.
For definiteness, and without losing in generality, let us work from now on with $j=1$. In this case
$P_1^\varphi = |B_1^\varphi\rangle \langle B_1^\varphi|$, with $|B_1^\varphi\rangle = \cos\varphi |00\rangle + \sin\varphi|11\rangle$.
If we set $\cos\varphi = \lambda_1$, $\sin\varphi = \lambda_2$, and remember that Alice is free to choose $\varphi$ in the range $[0,\pi/2]$,  
we readily see that $|B_1^\varphi\rangle=\lambda_1 |00\rangle + \lambda_2|11\rangle$ represents a Schmidt decomposition
of an \textit{arbitrary} two-qubit pure state $|\tilde{B}_1\rangle = a_{11}|u_1\rangle|v_1\rangle+a_{12}|u_1\rangle|v_2\rangle
+a_{21}|u_2\rangle|v_1\rangle+a_{22}|u_2\rangle|v_2\rangle$, where $|u_i\rangle$ and $|v_i\rangle$, $i=1,2$,
are any basis one can employ to expand the first and second qubits, respectively.
Thus, if we write the unitary transformation connecting these two states as $|B_1^\varphi\rangle=(U_{in} \otimes U^A) |\tilde{B}_j\rangle$,
and this can always de done \textit{locally} due to the properties of the Schmidt decomposition, Eq.~(\ref{Pjtilde}) becomes for $j=1$,
\begin{eqnarray}
\tilde{P}_1 &=& (U_{in}^\dagger \otimes (U^A)^\dagger)P_1^\varphi (U_{in} \otimes U^A) \nonumber \\
&=&(U_{in}^\dagger \otimes (U^A)^\dagger)|B_1^\varphi\rangle\langle B_1^\varphi| (U_{in} \otimes U^A) \nonumber \\
&=& |\tilde{B}_1\rangle\langle \tilde{B}_1|,
\label{Pjtilde2}
\end{eqnarray}
which shows that $\tilde{P}_1$ is an arbitrary projector. The same unitary operations above when applied to
$P_2^\varphi,P_3^\varphi$, and $P_4^\varphi$ generate the other three projectors $\tilde{P}_2,\tilde{P}_3,\tilde{P}_4$
that together with $\tilde{P}_1$ form a complete set of orthogonal projectors describing an arbitrary projective measurement.

We now move on to show that the classical limit proof given here also applies when Bob implements more general 
unitary operations on his qubit. The argument we use is similar to the one just developed above. It lies in the fact that 
$\langle \overline{F} \rangle_{\!_{\rho^{AB}}} \leq 2/3$ for \textit{any} separable state $\rho^{AB}$ and to using this property 
to conveniently express $\rho^{AB}$.

Similarly to what we did before, 
we rewrite Eq.~(\ref{noEntanglement}) as follows,
\begin{equation}
\tilde{\rho}^{AB} = \sum_{k=1}^n p_k \rho^A_k \otimes \tilde{\rho}^B_k,
\label{noEntanglementB}
\end{equation}
where $\tilde{\rho}^B_k=U^B\rho^B_k (U^B)^\dagger$ and $U^B$ is an 
arbitrary unitary operator acting on the Hilbert space of qubit $B$. 
The state describing the input qubit and $\tilde{\rho}^{AB}$ reads
\begin{equation}
\tilde{\rho} = U^B (\rho_{in} \otimes \rho^{AB} ) (U^B)^\dagger. 
\label{rhotildeB}
\end{equation}

After Alice's measurement $\tilde{\rho}$ changes to 
\begin{equation}
\tilde{\rho} \longrightarrow \frac{P_j^\varphi \tilde{\rho} P_j^\varphi}{\mbox{Tr}[P_j^\varphi \tilde{\rho}P_j^\varphi]} 
\end{equation}
and tracing out Alice's qubits we get Bob's state,
\begin{equation}
\tilde{\rho}_{\!_{B_j}} = \frac{\mbox{Tr}_{in,A}[P_j^\varphi \tilde{\rho} P_j^\varphi]}{\mbox{Tr}[P_j^\varphi \tilde{\rho}]}.
\label{bobjB}
\end{equation}
Inserting Eq.~(\ref{rhotildeB}) into (\ref{bobjB}) we obtain
\begin{equation}
\tilde{\rho}_{\!_{B_j}} = \frac{U^B\mbox{Tr}_{in,A}[P_j^\varphi \rho P_j^\varphi](U^B)^\dagger}
{Q_j(|\psi\rangle_{in})},
\label{bobj2B}
\end{equation}
where $\rho = \rho_{in}\otimes \rho^{AB}$ and $Q_j(|\psi\rangle_{in})=\mbox{Tr}[P_j^\varphi \rho]$.
After Bob implements the corresponding unitary operation $U_j$ on his qubit we arrive at the final output
state after a single run of the teleportation protocol,
\begin{eqnarray}
\rho_{\!_{B_j}} &=& \frac{U_jU^B\mbox{Tr}_{in,A}[P_j^\varphi \rho P_j^\varphi](U^B)^\dagger(U_j)^\dagger}
{Q_j(|\psi\rangle_{in})}, \nonumber \\ 
&=& \frac{U_jU^B\mbox{Tr}_{in,A}[P_j^\varphi \rho P_j^\varphi](U_jU^B)^\dagger}
{Q_j(|\psi\rangle_{in})}, \nonumber \\ 
&=& \frac{V_j\mbox{Tr}_{in,A}[P_j^\varphi \rho P_j^\varphi]V_j^\dagger}
{Q_j(|\psi\rangle_{in})}.
\label{bobj2BB}
\end{eqnarray}
Equation (\ref{bobj2BB}) is exactly Eq.~(\ref{twelve}) if we change $V_j=U_jU^B$ to $U_j$.
Also, $\mbox{Tr}[V_jV_k^\dagger] =\mbox{Tr}[U_jU^B(U_kU^B)^\dagger] = 
\mbox{Tr}[U_jU^B(U^B)^\dagger U_k^\dagger]=\mbox{Tr}[U_j U_k^\dagger]=2\delta_{jk}$.

We can repeat the previous arguments leading to Eq.~(\ref{bobj2BB}) without using explicitly the
fact that $\tilde{\rho} = U^B (\rho_{in} \otimes \rho^{AB} ) (U^B)^\dagger$. This gives
\begin{equation}
\rho_{\!_{B_j}}  =  \frac{U_j\mbox{Tr}_{in,A}[P_j^\varphi \tilde{\rho} P_j^\varphi]U_j^\dagger}
{Q_j(|\psi\rangle_{in})}.
\label{bobj2BBB}
\end{equation}
Now, since Eqs.~(\ref{bobj2BB}) and (\ref{bobj2BBB}) are equal, both representations of 
$\rho_{\!_{B_j}}$ when inserted into Eq.~(\ref{Fj}) will give the same expressions, which,
when employed to compute the deterministic average fidelity, as given by Eq.~(\ref{deterministicF}),
will furnish the same results: 
$\langle \overline{F} \rangle_{\rho^{AB},V_j} = \langle \overline{F} \rangle_{\tilde{\rho}^{AB},U_j}$.
Here the subscripts $V_j$ and $U_j$ remind us of which set of unitary operations one must use in the evaluations
of Eq.~(\ref{deterministicF}).
But we know that $\langle \overline{F} \rangle_{\tilde{\rho}^{AB},U_j}\leq 2/3$, since 
we already proved that the average fidelity for the deterministic protocol is upper bounded by $2/3$
for \textit{any} non-entangled state shared between Alice and Bob and when Bob uses the set $U_j$. 
Thus, we arrive at the desired result,
\begin{equation}
\langle \overline{F} \rangle_{\rho^{AB},V_j} \leq 2/3.
\end{equation}

\end{document}